\documentclass[12pt]{article}
\pdfoutput=1
\usepackage{eucal}
\let\mysub=\subset
\newcommand\bigsupset[1][1.19]{%
   \mathrel{\vcenter{\hbox{\scalebox{#1}{$\supset$}}}}}
\usepackage{graphicx}
\usepackage[dvipsnames]{xcolor}
\usepackage[hyphens]{url}
\usepackage{hyperref}
\hypersetup{
    colorlinks=false,
    linkbordercolor	=BlueViolet,
citebordercolor	=OliveGreen,
urlbordercolor=RoyalBlue
}

\usepackage{amsmath}
\usepackage{amssymb}

\DeclareMathAlphabet\mathrsfso      {U}{rsfso}{m}{n}

\numberwithin{equation}{section}
\usepackage{array}
\usepackage{mathtools}
\usepackage{dsfont}

\usepackage{tikz}\usetikzlibrary{matrix,fit,decorations.pathmorphing, positioning, shapes, backgrounds, decorations.markings}
\tikzset{snake it/.style={decorate, decoration=snake}}
\usepackage{varwidth}
\usepackage{enumerate}

\usepackage[margin=1in]{geometry}

\DeclareFontFamily{U}{solomos}{}
\DeclareErrorFont{U}{solomos}{m}{n}{10}
\DeclareFontShape{U}{solomos}{m}{n}{
  <-> s*[1.1]  gsolomos8r
}{}

\newcommand{\vkappa}{\text{\usefont{U}{solomos}{m}{n}\symbol{'153}}}

\usepackage{mathabx}
\usepackage{empheq}

\usepackage{setspace}

\usepackage[
    backend=bibtex,
    style=alphabetic,
maxbibnames=99,
giveninits=true,
minalphanames=1,
maxalphanames=3
  ]{biblatex}

\bibliography{SUSYCPNarxiv}

\newbibmacro{string+doi}[1]{%
  \iffieldundef{doi}{#1}{\href{http://dx.doi.org/\thefield{doi}}{#1}}}
\DeclareFieldFormat{title}{\usebibmacro{string+doi}{\mkbibemph{#1}}}
\DeclareFieldFormat[article]{title}{\usebibmacro{string+doi}{\mkbibquote{#1}}}

\usepackage{slashed}
\usepackage{upgreek}
\usepackage[titletoc]{appendix}

\usepackage{tabstackengine}
\fixTABwidth{T}

\newcommand{\dd}{\partial}
\newcommand{\bd}{\overline{\partial}}
\newcommand{\CP}{\mathsf{CP}}
\newcommand{\CC}{\mathsf{C}}
\newcommand{\Phis}{\mathsf{\Phi}}
\newcommand{\Ms}{\mathsf{M}}

\renewcommand{\bar}{\overline}
\renewcommand{\tilde}{\widetilde}

\newcommand{\GL}{\mathbf{GL}}
\newcommand{\SL}{\mathbf{SL}}
\newcommand{\PSL}{\mathbf{PSL}}
\newcommand{\G}{\mathbf{G}}

\newcommand{\bea}{\begin{equation}}
\newcommand{\eea}{\end{equation}}
\newcommand{\bear}{\begin{eqnarray}}
\newcommand{\eear}{\end{eqnarray}}
\newcommand{\bearr}{\begin{eqnarray*}}
\newcommand{\eearr}{\end{eqnarray*}}

\newcommand{\appendixnumberline}[1]{Appendix #1.\space}

\let\oldappendix\appendix
\makeatletter
\renewcommand{\appendix}{%
  \addtocontents{toc}{\let\protect\numberline\protect\appendixnumberline}%
  \renewcommand{\@seccntformat}[1]{\large\bfseries Appendix . }%
  \oldappendix
}
\makeatother

\usepackage{mdframed}

\newmdenv[
  topline=false,
  bottomline=false,
  linewidth=2pt,
  skipabove=\topsep,
  skipbelow=\topsep
]{siderulesright}

\makeatletter
\renewcommand{\@seccntformat}[1]{\csname the#1\endcsname.\quad}
\makeatother

\usepackage{setspace}
\usepackage{xpatch}

\makeatletter
\renewcommand{\@chap@pppage}{%
  \clear@ppage
  \thispagestyle{plain}%
  \if@twocolumn\onecolumn\@tempswatrue\else\@tempswafalse\fi
  \null\vfil
  \markboth{}{}%
  {\centering
   \interlinepenalty \@M
   \normalfont
   \MakeUppercase \appendixpagename\par}%
  \if@dotoc@pp
    \addappheadtotoc
  \fi
  \vfil\newpage
  \if@twoside
    \if@openright
      \null
      \thispagestyle{empty}%
      \newpage
    \fi
  \fi
  \if@tempswa
    \twocolumn
  \fi
}
\makeatother

\renewbibmacro{in:}{%
  \ifentrytype{article}
    {}
    {\printtext{\bibstring{in}\intitlepunct}}}

\usepackage{tocloft}

\addtocontents{toc}{%
    \protect}

\let \savenumberline \numberline
\def \numberline#1{\savenumberline{#1.}}

\usepackage{etoolbox}
\patchcmd{\tableofcontents}{\@starttoc}{\vspace{-0.3cm}\@starttoc}{}{}

\usepackage{titlesec}

\titleformat*{\section}{\large\bfseries\sf}
\titleformat*{\subsection}{\normalsize\bfseries}
\titleformat*{\subsubsection}{\normalsize\bfseries}
\titleformat*{\paragraph}{\large\bfseries}
\titleformat*{\subparagraph}{\large\bfseries}
\titlespacing{\author}{-5pt}{-5pt}{-5pt}[-5pt]

\makeatletter
\renewcommand\subsubsection{\@startsection{subsubsection}{3}{\z@}%
                                     {-3.25ex\@plus -1ex \@minus -.2ex}%
                                     {-1.5ex \@plus -.2ex}
                                     {\normalfont\normalsize\bfseries\sf}}
\renewcommand\subsection{\@startsection{subsection}{3}{\z@}%
                                     {-3.25ex\@plus -1ex \@minus -.2ex}%
                                     {-1.5ex \@plus -.2ex}
                                     {\normalfont\normalsize\bfseries\sf}}                                     
\makeatother

   \usepackage{stmaryrd}

   \usepackage{enumitem}
   
\definecolor{mygreen1}{RGB}{0, 204, 102}

\begin{document}

\title{\vspace{-1.0cm} \sf The $\CP^{n-1}$-model with fermions: a new look}
\author{Dmitri Bykov$^{1,2, 3}$\footnote{Emails:
bykov@mpp.mpg.de, bykov@mi-ras.ru, dmitri.v.bykov@gmail.com}
\\  \vspace{-0.3cm}  \\
{\footnotesize $^1$ Max-Planck-Institut f\"ur Physik, F\"ohringer Ring 6, 80805 Munich, Germany} \vspace{-0.2cm}\\ \vspace{-0.2cm}
{\footnotesize $^2$ Arnold Sommerfeld Center for Theoretical Physics, Theresienstrasse 37, 80333 Munich, Germany}  
\\ 
\vspace{-0.2cm}
{\footnotesize $^3$ Steklov
Mathematical Institute of Russ. Acad. Sci., Gubkina str. 8, 119991 Moscow, Russia \;}
}

\date{}

\begin{flushright}    
  {\small
    MPP-2020-166 \\
    LMU-ASC-35/20 
  }
\end{flushright}
\vspace{-0.3cm}
{\let\newpage\relax\maketitle}

\maketitle

\vspace{-0.7cm}
\begin{siderulesright}\small
We elaborate the formulation of the $\CP^{n-1}$ sigma model with fermions as a gauged Gross-Neveu model. This approach allows to identify the super phase space of the model as a supersymplectic quotient. Potential chiral gauge anomalies are shown to receive contributions from bosons and fermions alike and are related to properties of this phase space. Along the way we demonstrate that the worldsheet supersymmetric model is a supersymplectic quotient of a model with target space supersymmetry. Possible generalizations to other quiver supervarieties are briefly discussed.
\end{siderulesright}

\vspace{-0.3cm}
\tableofcontents

\section{\sf Introduction and main results}

The goal of the present paper is to take a new look at the well-known $\CP^{n-1}$ {$\sigma$-model}~\cite{Cremmer, DAdda1, AddaSUSY} on a two-dimensional Euclidean worldsheet, particularly in the case when it is coupled to fermions in various ways. Since $\CP^{n-1}$ is a symmetric space, it has long been known that the bosonic model is classically integrable~\cite{ZM} in the sense of the inverse scattering method. As usual, integrability implies the existence of an infinite number of conserved charges in involution. Only a finite subset of these charges is related to the `obvious' global symmetries of $\CP^{n-1}$, and the rest of the charges may be found, for instance, from the celebrated zero-curvature (Lax pair) representation for the equations of motion (e.o.m.). In fact, from the latter one can derive both local charges in involution~\cite{PolyakovCharges} as well as non-local charges~\cite{LuscherPohl, LuscherNonlocal}) generating an infinite-dimensional `quantum algebra' (in infinite volume)~\cite{Bernard1, Bernard2}, see~\cite{Loebbert} for a review. For complex symmetric spaces, such as $\CP^{n-1}$ or Grassmannians, these charges are anomalous at the quantum level, which therefore spoils integrability. For the local charges this was shown in~\cite{Goldschmidt}, and for the non-local charges in~\cite{AbdallaAnomaly}. In~\cite{AbdallaCPN} it was observed that the latter anomalies are canceled, if the $\sigma$-model is coupled to fermions in suitable ways, for example minimally or supersymmetrically (see also~\cite{AbdallaBook} for a review of these developments).
More general theories with fermions were considered in~\cite{AbdallaFermions}, where it was noticed that the mechanism by which the anomaly in the nonlocal charge is cancelled is related to the well-known chiral anomalies. In that paper the authors remark that ``\emph{The deeper reasons for this miraculous anomaly cancellation, however, remain obscure, and the question certainly deserves further investigation}''. The present paper is a step towards a geometric explanation of the anomaly cancellation.

\vspace{0.3cm}\noindent
Our starting point is a novel presentation of the  $\CP^{n-1}$-model as a gauged Gross-Neveu model proposed in~\cite{BykovGN}.
This presentation is essentially a composition of the $\beta\gamma$-system approach of~\cite{CYa} and the gauged linear $\sigma$-models~(GLSM) for flags developed in~\cite{BykovGLSM1, BykovGLSM2, BykovAnom}. This approach may be easily extended to incorporate fermions (in various ways, including supersymmetric couplings\footnote{General supersymmetric $\beta\gamma$-systems have been recently studied in~\cite{Rocek2020}.}), which then leads to boson-fermion `superpositions' of chiral Gross-Neveu models. In particular, we will show that this framework  provides a new way of constructing models with worldsheet supersymmetry by starting from models with target space supersymmetry and gauging (super)subgroups of their global symmetry groups\footnote{In the present paper we will be dealing with the $\CP^{n-1}$-model that admits $\mathcal{N}=(2, 2)$ SUSY, but the method seems to be inherently applicable to models with $\mathcal{N}=(0, 2)$ SUSY (see~\cite{Distler} for a concise review of the latter).}. The following fact will be of foremost importance: if one is to deal with a projective target space (such as most of the target spaces we are interested in: $\CP^{n-1}$, Grassmannians, flag manifolds), one needs to gauge part of the chiral symmetry, under which both fermions \emph{and bosons} are charged in our approach. The problem is that this symmetry is typically anomalous, and the gauging can be performed quantum-mechanically only if the anomaly is cancelled. As we shall see, the condition of anomaly cancellation is a simple constraint on the representations of the supergroups, in which the matter fields transform. In the known examples the cancellation of such chiral anomalies also implies the cancellation of the anomalies in the Yangian (i.e. in the integrability charges). Although we leave the full construction of the quantum theory for future work, we believe this to be the general case.

\vspace{0.3cm}\noindent
Apart from the analysis of anomalies, one of the goals of this paper is to provide a differential-geometric setup for a rather wide class of integrable $\sigma$-models incorporating fermions. The integrability of the original purely fermionic Gross-Neveu model~\cite{GrossNeveu} was observed shortly after it was put forward~\cite{Dashen}, and soon it was realized that it is related to the integrability of bosonic models formulated in terms of the fermion bilinears~\cite{NeveuPapa}. In the present paper we interpret these bilinears as moment maps for the action of various symmetry groups  on complex symplectic manifolds. We also demonstrate that the interactions in the models may be, quite universally, written as products $\mathrm{Tr}(\mu\bar{\mu})$ of the moment map with its complex conjugate. The same approach may be used for the analysis of rather general `quiver supervarieties' satisfying anomaly cancellation conditions. 

\vspace{0.3cm}\noindent
The structure of the paper is naturally entangled with the two pieces of data that determine the $\sigma$-models in question: the `phase space' of the model and a `Hamiltonian', or interaction term. In section~\ref{phasespacesec} we discuss the super phase spaces of the $\CP^{n-1}$-model with fermions. In section~\ref{interactionsec} we introduce the interactions and prove the supersymmetry of the model in a special case. The condition for the cancellation of chiral gauge anomalies is explained in section~\ref{anomalysec}, followed by an explanation of the role of the gauge fields in these models in section~\ref{QMsec} on the example of a quantum-mechanical reduction. In section~\ref{quiversec} we present a general differential-geometric setup involving `quiver supervarieties', to which the methods of the present paper apply.

\vspace{0.3cm}\noindent
\emph{Notation.} We will assume that the worldsheet is the complex plane $\CC$, with coordinates~$z, \bar{z}$. Derivatives with respect to these coordinates will be denoted $\dd:=\dd_z$ and $\bd:=\dd_{\bar{z}}$. Similar notation $\mathrsfso{D}, \bar{\mathrsfso{D}}$ is adopted for covariant derivatives. All Lie groups and algebras are assumed to be defined over complex numbers, unless a real form such as $SU(n)$ is explicitly referred to.

\section{The $\CP^{n-1}$-model with fermions: phase space}\label{phasespacesec}
We pass to the definition of the $\sigma$-model with target space $\Ms=\CP^{n-1}$, first in the purely bosonic case. As already mentioned, we will be using the formulation of the model as a gauged Gross-Neveu model~\cite{BykovGN}, which may be thought of as a coupling of two $\beta\gamma$-systems~\cite{CYa} in a GLSM formalism of~\cite{BykovGLSM1, BykovGLSM2, BykovNilp}. We will start by writing the Lagrangian in a slightly more general form than actually needed for our purposes in the present paper, to emphasize that the methods explained here may as well be applied to the case of trigonometrically/elliptically deformed models. The Lagrangian is 
\bear\label{lagr00}
&&\mathrsfso{L}=V\cdot  \bar{\mathrsfso{D}} U+\bar{U}\cdot  \mathrsfso{D} \bar{V}+\mathrm{Tr}\left(r_s(U V) (U V)^\dagger\right)\,,\quad\quad\\ \nonumber&&\textrm{where}\quad\quad \bar{\mathrsfso{D}}U=\bd U+i \,U \bar{\mathcal{A}}\,.
\eear
Here $U$ and $V$ are $n$-component column- and row-vectors respectively, $\bar{\mathcal{A}}$ is an auxiliary gauge field and $r_s$ is the classical $r$-matrix (rational, trigonometric or elliptic) depending on the spectral parameter `$s$' and satisfying the classical Yang-Baxter equation~\cite{BelavinDrinfeld}. For the sake of completeness let us point out that this deformation is (in general) not the same as the $\eta$-deformation~\cite{Klimcik, DMVq} that relies on the symmetric space structure of the target manifold. The $\eta$-deformation of $\CP^{n-1}$ was studied in detail in~\cite{Litvinov19, FateevDuality19, Demulder, BykovLust}, in particular in the last paper the relation between the two deformations is discussed. In what follows we will focus on the \emph{rational} case, in which the $r_s$-matrix is proportional to the identity operator: $r_s=\mathrm{Id}$. The discussion of anomalies in sec.~\ref{anomalysec} below would not be altered by the deformation, but we leave a detailed study of deformations for future work.

\vspace{0.3cm}\noindent
In the present section we consider the possible ways of including fermions in the bosonic model written above. To do this, first we will ignore the interaction term in~(\ref{lagr00}) and concentrate on the kinetic term instead. The key observation is that the kinetic term is naturally defined in terms of a Liouville one-form $\uptheta$ corresponding to a certain complex symplectic form $\Upomega=d\uptheta$. As the gauge field in~(\ref{lagr00}) suggests, this symplectic form is the symplectic reduction of the standard form $\Upomega^{(0)}=\sum\limits_{i=1}^n \,dV_i\wedge  dU_i$ on $\mathsf{T}^\ast\CC^n$ under the $\CC^\times$-action that scales the $(U, V)$-coordinates: $U\to \uprho U, V\to \uprho^{-1} V$. In order to include fermions, one should pass over to symplectic supervarieties and superquotients thereof. Below we elaborate the two most notable phase spaces that arise in this way\footnote{We will see later that one can have several different theories for a given phase space, so that the `minimal fermions phase space' or `supersymmetric phase space' is just a convenient way to label those supermanifolds.}: that of `minimal  fermions' in sec.~\ref{minphasesec} and that of the supersymmetric theory in sec.~\ref{susyphasesec}.

\vspace{0.1cm}
\subsection[Minimal fermions phase space]{Minimal fermions phase space.}\label{minphasesec} Ignoring interactions, we may write the kinetic part of the $\CP^{n-1}$ Lagrangian with an additional Dirac fermion $\Theta$ as follows:
\bea\label{lagrkin1}
\mathrsfso{L}_{\mathrm{min}}=\bar{\Psi} \slashed{\mathrsfso{D}}\Psi+\bar{\Theta} \slashed{\mathrsfso{D}}\Theta\,,
\eea
where $\Psi$ is a bosonic spinor $\Psi=\begin{pmatrix}U\\ \bar{V}\end{pmatrix}$ and $\Theta$ a fermionic one: $\Theta=\begin{pmatrix}C\\ \bar{B}\end{pmatrix}$. In the above formula $\mathrsfso{D}$ is a $\CC^\times$-covariant derivative, where the action of $\CC^\times$ is as follows: $U\to \uprho U, V\to \uprho^{-1} V, C\to\uprho C, B\to  \uprho^{-1}B $. The first term in~(\ref{lagrkin1}) coincides with the kinetic term in the Lagrangian~(\ref{lagr00}). Viewing $\Psi$ and $\Theta$ as coordinates on the superspace $\CC^{n|n}$, we conclude that the phase space of the model~(\ref{lagrkin1}) is
\begin{empheq}[box=\fbox]{align}
\label{phaseminferm}
\Phis_{\textrm{min}}=(\sf{T^\ast} \CC^{n|n})\!\sslash\! \CC^\times\,.
\end{empheq}

\noindent
The notation $\sslash$ means `complex symplectic quotient'. The phase space $\Phis_{\textrm{min}}$ contains a dense open subspace $\Phis_{\textrm{min}}^{(st)}$ (`$st$' for `stable'), where $U\neq 0$, such that $\Phis_{\textrm{min}}^{(st)}=\sf{T^\ast M}_{\textrm{min}}$, and $\Ms_{\textrm{min}}$ is the configuration space
\begin{empheq}[box=\fbox]{align}
\hspace{1em}
\label{confminferm}
\Ms_{\textrm{min}}=\CC^{n|n} / \CC^\times\,=\CP^{n-1|n}\,.
\end{empheq}
The quotient $/$ should be understood as a geometric invariant theory (GIT)-quotient (for a more detailed discussion of such subtleties we refer the reader to~\cite{NakajimaReview}).  The space~(\ref{confminferm}) is the same as the total space of the vector bundle $\mathsf{\Pi V}$, where $\mathsf{\Pi}$ means that the fibers are fermionic (Grassmann), and $\mathsf{V}=\mathcal{O}(1)\oplus \cdots \oplus \mathcal{O}(1)$. This is a super-Calabi-Yau manifold, the holomorphic nowhere vanishing top form being (we use the inhomogeneous coordinates,~$U_n=1$)
\bea
\Omega_{\mathrm{min}}=dU_1\wedge \cdots \wedge dU_{n-1}\wedge d C_1 \wedge \cdots \wedge dC_n\,.
\eea
This is a fermionic and higher-$n$ analogue of the conifold $\mathcal{O}(-1)\oplus \mathcal{O}(-1)\to \CP^1$.

\vspace{0.5cm}
\subsection[Supersymmetric phase space]{Supersymmetric phase space.}\label{susyphasesec} For the original definitions of the supersymmetric $\CP^{n-1}$-model we refer to~\cite{WittenCP1} (the case of $\CP^1$) and~\cite{Cremmer, AddaSUSY} (arbitrary $n$), see also~\cite[Chapter 15]{Mirror} for a more modern treatment. Our approach is based on the following observation:

\vspace{0.3cm}\noindent
\begin{center}
\noindent\fbox{%
    \parbox{14cm}{\vspace{-0.3cm}
    \begin{center}
\emph{The worldsheet supersymmetric $\CP^{n-1}$-model is a gauged version of a model with target space $\mathbf{GL}(1|1)$ supersymmetry.}
\end{center}
\vspace{-0.3cm}
}}
\end{center}

\vspace{0.3cm}\noindent
The general strategy of proving the statement is as follows:  we start with the ungauged $\beta\gamma$-system with phase space $\sf{T^\ast} \CC^n$, impose worldsheet SUSY, gauge the $\CC^\times$-symmetry to obtain projective space and, finally, introduce interactions in sec.~\ref{SUSYsec} below.

\vspace{0.3cm}\noindent
The $\beta\gamma$-system with phase space $\sf{T^\ast} \CC^n$ reads:
\bea
\mathrsfso{L}_{\beta\gamma}=V\cdot \bd U+\bar{U}\cdot\dd \bar{V}  \,.
\eea
In order to supersymmetrize the model, we add a fermionic piece symmetrically:
\bea\label{freelagr}
\widetilde{\mathrsfso{L}_{\beta\gamma}}=(V\cdot \bd U+\bar{U}\cdot\dd \bar{V})+(B\cdot \bd C+\bar{C}\cdot \dd\bar{B})\,.
\eea
Here $C_i, B_i$ ($i=1, \ldots, n$) are the fermionic variables\footnote{The Lagrangian may also be written as a sum of two Dirac pieces: $\widetilde{\mathrsfso{L}_{\beta\gamma}}=\bar{\Psi} \slashed{\dd}\Psi+\bar{\Theta} \slashed{\dd}\Theta$, where $\Psi$~is a bosonic spinor $\Psi=\begin{pmatrix}U\\ \bar{V}\end{pmatrix}$ and $\Theta$ a fermionic one: $\Theta=\begin{pmatrix}C\\ \bar{B}\end{pmatrix}$.}. The holomorphic piece (depending on $U, V, B, C$) is invariant w.r.t. both target space and worldsheet SUSY transformations. We pass to the discussion of these.

\subsubsection{Target-space supersymmetry.} The full target space symmetry group of~(\ref{freelagr}) is $\GL(n|n)$. Indeed, in terms of the doublets
\bea\label{UVdoublet}
   \mathrsfso{U}:=\begin{pmatrix} 
      U  \\
      C  \\
   \end{pmatrix},\quad\quad \mathrsfso{V}:=\begin{pmatrix} \,V & B\,\,\end{pmatrix}
\eea
the Lagrangian is $\mathrsfso{L}^{SUSY}_{\beta\gamma}=\mathrsfso{V}\cdot \bd\mathrsfso{U}+ \bar{\mathrsfso{U}}\cdot \dd\bar{\mathrsfso{V}}$, so that it is manifestly invariant under
\bea
\mathrsfso{U}\to g\cdot \mathrsfso{U},\quad\quad \mathrsfso{V}\to \mathrsfso{V}\cdot g^{-1},\quad\quad g\in \GL(n|n)\,.
\eea
In most of the models of interest, however, we will be introducing interactions that are invariant only under the diagonal subgroup
\bea
\G_0:=\GL(1|1)\mysub \GL(n|n)
\eea
In what follows it will be convenient to use an explicit parametrization of the matrix~${g\in \G_0}$:
\bea\label{gl11mat}
g=\begin{pmatrix} 
      \uplambda\,\uprho & \upchi  \\
      \upxi &  \uplambda
   \end{pmatrix},\quad\quad \mathrm{SDet}(g)=\uprho-{\upchi \upxi\over \uplambda^2}\,.
\eea

\subsubsection{Worldsheet supersymmetry.} The Lagrangian~(\ref{freelagr}) is in addition invariant\footnote{Up to a total derivative, as is usual for supersymmetry.} w.r.t. the  worldsheet supersymmetry transformations. The generators of the (right-moving/holomorphic) transformations are customarily denoted in the literature as $Q_+, \bar{Q}_+$, and the variation of the fields is obtained by acting on them with the operator $\delta=\upepsilon_1 Q_++\upepsilon_2 \bar{Q}_+$ ($\upepsilon_{1, 2}$ are complex Grassmann variables):
\bea\label{elemSUSY}
\delta U=\upepsilon_1 \,C,\quad\quad \delta B=- \upepsilon_1 \,V,\quad\quad \delta C= -\upepsilon_2 \,\dd U,\quad\quad \delta V=\upepsilon_2 \,\dd B\,.
\eea
These elementary transformations have been discussed in~\cite{Kapustin, Policastro}\footnote{In~\cite{Kapustin} the $\sigma$-model was shown to be supersymmetric w.r.t. the transformations~(\ref{elemSUSY}) in the `infinite-volume' limit of the target space. In our formulation, as we will see, the gauged analogues~(\ref{gaugedSUSY}) of these transformations are exact.}. It follows that the charges satisfy the $(0, 2)$ supersymmetry algebra:~$Q_+^2=\bar{Q}_+^2=0, \{Q_+, \bar{Q}_+\}=\dd$. The anti-holomorphic piece in the Lagrangian is analogously invariant w.r.t. the left-moving SUSY transformations. Notice that from the perspective we adopt here $Q_+$ is actually one of the generators of the target-space symmetry algebra $\mathfrak{gl}(1|1)$.

\vspace{0.3cm}\noindent
In order to pass to the case of $\CP^{n-1}$ we replace the derivatives by covariant ones, i.e.
\bear\label{cpfreelagr}
&&\mathrsfso{L}_{\CP}= (V\cdot \bar{\mathrsfso{D}} U+\bar{U}\cdot\mathrsfso{D} \bar{V})+(B\cdot \bar{\mathrsfso{D}} C+\bar{C}\cdot \mathrsfso{D}\bar{B})\\ \nonumber
&&\bar{\mathrsfso{D}} =\bd +i\,\bar{\mathcal{A}}\,.
\eear
We have gauged a $\CC^\times \mysub \GL(1|1)$ subgroup corresponding to $\uprho=1, \upxi=\upchi=0$ in~(\ref{gl11mat}). Varying the above Lagrangian w.r.t. the gauge field produces a constraint
\bea
\label{mm1}
V \cdot U+B \cdot C=0\,.
\eea

\noindent
The supersymmetry transformations now take the form
\bea\label{gaugedSUSY}
\delta U=\upepsilon_1 \,C,\quad\quad \delta B=- \upepsilon_1 \,V,\quad\quad \delta C= -\upepsilon_2 \,\mathrsfso{D} U,\quad\quad \delta V=\upepsilon_2 \,\mathrsfso{D} B\,.
\eea
In this case $Q_+^2=\bar{Q}_+^2=0, \{Q_+, \bar{Q}_+\}=\mathrsfso{D}$. The variation of the Lagrangian is
\bear
&&\delta \mathrsfso{L}_{\CP}=\upepsilon_2 \,\mathrsfso{D} B\cdot \bar{\mathrsfso{D}} U+\upepsilon_2\,B\cdot \bar{\mathrsfso{D}} \mathrsfso{D} U=\\ \nonumber
&&=\upepsilon_2 \,\dd( B\cdot \bar{\mathrsfso{D}} U)+\upepsilon_2\, F_{z\bar{z}}\,(B \cdot U),\quad\quad F_{z\bar{z}}=[\bar{\mathrsfso{D}}, \mathrsfso{D}]\,.
\eear
The variation is a total derivative (and hence the action is invariant) if and only if
\bea
\label{mm2}
B \cdot U=0\,.
\eea
This condition is invariant under supersymmetry:
\bea
\delta(B \cdot U)=-\upepsilon_1(V \cdot U+B \cdot C)=0\,.
\eea
The variation is proportional to the constraint~(\ref{mm1}) and therefore vanishes on-shell.

\subsubsection{The super phase space.}\label{SUSYphasesec}
If the condition~(\ref{mm2}) is satisfied, the Lagrangian~(\ref{cpfreelagr}) is invariant under an additional \emph{local} symmetry
\bea\label{locsymm}
\delta_{\mathrm{loc}} C=\updelta(z, \bar{z}) \,U,\quad\quad \delta_{\mathrm{loc}} V= \updelta(z, \bar{z}) \,B\,,
\eea
where $\updelta(z, \bar{z}) $ is a local Grassmann variable. Recalling the $\CC^\times$ gauge group, we observe that we are actually gauging a subgroup $\G_{\triangle}\mysub \GL(1|1)$ comprising matrices of the form
\bea\label{Gtriang}
\G_{\triangle}:=\left\{\quad g\in \SL(1|1)\,:\quad g=\begin{pmatrix} 
      \uplambda & 0  \\
      \upxi &  \uplambda
   \end{pmatrix}\quad \right\}
\eea
The constraints~(\ref{mm1}), (\ref{mm2}) should be seen as the moment map constraints, and the phase space itself should be viewed as the supersymplectic reduction
\begin{empheq}[box=\fbox]{align}
\label{phaseSUSY}
\Phis_{\textrm{SUSY}}=(\mathsf{T^\ast} \CC^{n|n})\!\sslash\! \G_{\triangle}\,.
\end{empheq}

\noindent
An important point is that gauging breaks the global symmetry $\GL(n|n)$ of the ungauged model down to\footnote{$N_{\GL(n|n)}(\G_\triangle)$ is the normalizer of $\G_\triangle$ in $\GL(n|n)$.}
\bea\label{normalizer}
\frac{N_{\GL(n|n)}(\G_\triangle)}{\G_\triangle}\simeq (\CC^{\times}\!\!\times\PSL(n))\ltimes \mathsf{\Pi} (\mathfrak{sl}_n)=\left\{\begin{pmatrix} 
      \Uplambda & 0  \\
      0 &  \upsigma\,\Uplambda
   \end{pmatrix}\cdot \begin{pmatrix} 
      \mathbf{1}_n & 0  \\
      \Upxi &  \mathbf{1}_n
   \end{pmatrix}\mysub \GL(n|n)\right\} \,,
\eea
where $\upsigma \in \CC^\times$, $\PSL(n)$ (represented by $\Uplambda$) is embedded in $\GL(n|n)$ diagonally, $\mathsf{\Pi} (\mathfrak{sl}_n)$ (represented by $\Upxi$) are fermionic traceless $n\times n$-matrices\footnote{We view $\mathsf{\Pi} (\mathfrak{sl}_n)$ as a vector space. As a Lie algebra it is simply anti-commutative.} in the adjoint representation of the bosonic $\PSL(n)$ and are scaled by $\CC^\times$. In fact, in most applications the interactions will only be invariant under the subgroup $\GL(1|1)\times \G_B$ of $\GL(n|n)$, where $\G_B$ is a bosonic group, and in this case the gauging breaks the target-space supersymmetry group $\GL(1|1)$ down to $\CC^\times$. For example, such is the worldsheet supersymmetric $\CP^{n-1}$-model, where one does not expect any residual target space supersymmetry. In that case $\CC^{\times}$ is the $R$-symmetry group containing both vectorial and axial transformations\footnote{See section~\ref{anomalysec} below for an explanation of how chiral symmetry acts in Euclidean signature.} (but only the vectorial $U(1)\mysub \CC^{\times}$ is non-anomalous~\cite[Chapter~15]{Mirror}). 

\vspace{0.3cm}\noindent
Just as in the case~(\ref{confminferm}) of minimal fermions, we may identify the configuration space, if we restrict to the stable set $\Phis_{\textrm{SUSY}}^{(st)}\mysub \Phis_{\textrm{SUSY}}$ defined by the requirement $U\neq 0$. The action of $\G_\triangle$ on the coordinates $\mathrsfso{U}$ is $\mathrsfso{U}\to g\cdot \mathrsfso{U}$, where $g$ is of the form~(\ref{Gtriang}), which in components is $U\to \uplambda\, U, C\to \uplambda\, C+\upxi \,U$. If one ignores $C$, the quotient w.r.t. the $\CC^\times$ action with parameter $\uplambda$ simply leads to $\CP^{n-1}$. The role of $C$ is that it describes a certain vector bundle over that projective space. Taking the quotient by multiples of $U$ means one has a quotient bundle $\CC^n / \mathcal{O}(-1)$, where $\CC^n$ is the trivial bundle. An additional multiplication by $\uplambda$ means that one in fact has the bundle $\mathsf{V}=\mathcal{O}(1)\otimes (\CC^n / \mathcal{O}(-1))$, which is the tangent bundle $\mathsf{V}=\mathsf{T} \CP^{n-1}$. The configuration space is
\begin{empheq}[box=\fbox]{align}
\label{confSUSY}
\Ms_{\textrm{SUSY}}=\mathsf{\Pi(T} \CP^{n-1})\,.
\end{empheq}

\noindent
Again, due to the fact that the fibers are fermionic $\Ms_{\textrm{SUSY}}$ is super-Calabi-Yau (a fermionic analogue of the cotangent bundle $\mathsf{T^\ast}\CP^{n-1}$). A recent discussion of the properties of the spaces~(\ref{confminferm}) and~(\ref{confSUSY}) in the case $n=2$ may be found in~\cite{Noja}.

\section{Interactions}\label{interactionsec}
Next we come to the description of interactions. In particular, we would like to couple the holomorphic (with coordinates $U, V$) and anti-holomorphic ($\bar{U}, \bar{V}$) $\beta\gamma$-systems, so as to obtain the more conventional $\sigma$-models. This may be done rather beautifully by a coupling of the form
\bea\label{coupling}
\vkappa\, \mathrm{Tr}(\mu\,\bar{\mu})\,,
\eea
where $\mu$ is the moment map for the symplectic action of a group $\G$ on the phase space $\Phi$ of the model and $\vkappa$ is the coupling constant. As we will now explain, different choices of such action (and of the group $\G$ itself) will lead to different models. For simplicity we will always assume that $\G\bigsupset \PSL(n)$, although one could as well consider smaller symmetry groups. One should bear in mind that the coupling~(\ref{coupling}) breaks the complex symmetry group $\G$ down to its unitary subgroup.

\vspace{0.3cm}\noindent
Our first examples will refer to the `minimal' phase space $\Phi_{\mathrm{min}}$. The group of its symplectic automorphisms is $\PSL(n|n)$.

\vspace{0.5cm}
\subsection[The $\CP^{n-1|n}$ model]{The $\CP^{n-1|n}$ model.}

Choosing $\G=\PSL(n|n)$, we obtain a model with explicit $\mathbf{PSU}(n|n)$-symmetry (the unitary subgroup of $\G$ arising because of the interaction term~(\ref{coupling})). This is a sigma model with target space $\CP^{n-1|n}$ that has been studied in~\cite{ReadSaleur, SaleurSchomerus}. The most notable case -- $\CP^{3|4}$ -- related to the so-called twistor string theory, was thoroughly discussed in~\cite{WittenTwistor}. In terms of the doublets
\bea\label{doublets}
   \mathrsfso{U}:=\begin{pmatrix} 
      U  \\
      C  \\
   \end{pmatrix},\quad\quad \mathrsfso{V}:=\begin{pmatrix} \,V & B\,\, \end{pmatrix}
\eea
introduced in~(\ref{UVdoublet}) the moment map is, in this case,
\bea
\mu_{\PSL(n|n)}=\mathrsfso{U}\otimes \mathrsfso{V}\,.
\eea
We do not subtract the trace part $\mathrsfso{V} \cdot \mathrsfso{U}$, as it is assumed to vanish as a consequence of the constraint imposed by the $\CC^\times$ gauge field of the model.

\vspace{0.5cm}
\subsection[Minimal and non-minimal fermions]{Minimal and non-minimal fermions.}

The bosonic subgroup of  $\PSL(n|n)$ is $\SL(n)\times \SL(n)$, where the two factors act respectively on $\CC^{n|0}$ and $\CC^{0|n}$. The corresponding moment maps are\footnote{The signs in $\mu_{\mathrm{BC}}$ are due to the anti-commutativity of the fermions.} 
\bear
&&\mu_{\mathrm{UV}}=U\otimes V-{(V\cdot U)\over n}\,\mathbf{1}_n\,,\\
&&\mu_{\mathrm{BC}}=-C\otimes B-{(B\cdot C)\over n}\,\mathbf{1}_n\,.
\eear
Although we do not set the goal to classify all subgroups $\G \mysub \PSL(n|n)$ that lead to interesting models, some options seem especially natural:
\begin{itemize}[leftmargin=*]
\item $\G=\SL(n)\times \mathbf{1}$. The resulting model is that of $\CP^{n-1}$ with minimally coupled fermions (in the sense that the fermions do not enter the interaction terms~(\ref{coupling})).
\item $\G=\mathbf{1} \times \SL(n)$. This produces a model with `minimally coupled bosons'. In the ungauged case this is the original fermionic chiral Gross-Neveu model.
\item Diagonal and anti-diagonal embeddings $i: \G= \SL(n) \hookrightarrow \SL(n)\times \SL(n)$, where $i(g)=(g, g)$ for the diagonal embedding and $i(g)=(g, (g^{-1})^{\mathrm{T}})$ for the anti-diagonal~one. The corresponding moment maps are
\bear
&&\mu_{\mathrm{diag}}=\mu_{\mathrm{UV}}+\mu_{\mathrm{BC}}\\
&&\mu_{\mathrm{anti-diag}}=\mu_{\mathrm{UV}}-(\mu_{\mathrm{BC}})^{\mathrm{T}}\,.
\eear
\item $\G=\SL(n)\times \SL(n)$. This gives a completely symmetric coupling of a bosonic chiral Gross-Neveu model to a fermionic one, the coupling being mediated by a gauge field.
\end{itemize}
The constraint induced by the gauge fields is, in all cases, $\mathrsfso{V} \cdot \mathrsfso{U}=V\cdot U+B\cdot C=0$.

\vspace{0.5cm}
\subsection[Supersymmetric model]{Supersymmetric model.}\label{SUSYsec}

In the previous subsections we considered the `minimal' phase space $\Phi_{\mathrm{min}}$. Now we come to the discussion of the `supersymmetric' phase space~$\Phi_{\mathrm{SUSY}}$. As explained in section~\ref{SUSYphasesec}, its symplectomorphism group is $(\CC^{\times}\!\!\times\PSL(n))\ltimes \mathsf{\Pi} (\mathfrak{sl}_n)$. We will choose the subgroup $\G=\PSL(n)$ as the symmetry group and prove that this leads to the standard supersymmetric $\sigma$-model. The remaining $\CC^{\times}$-invariance is the classical $R$-symmetry of the supersymmetric theory.

\vspace{0.3cm}\noindent
The moment map for the action of $\mathfrak{gl}_n$ diagonally embedded in $\mathfrak{gl}_{n|n}$ is 
\bea
\mu:=\mu_{\mathrm{diag}}=U\otimes V-C\otimes B\,.
\eea
This is also the most general $\GL(1|1)$-invariant combination of the holomorphic variables and may as well be viewed as the $z$-component of the Noether current of the model~(\ref{cpfreelagr}) (as well as of the full interacting model~(\ref{SUSYlagr}) below) corresponding to the $\mathfrak{gl}_n$-symmetry. Note that $\mathrm{Tr}(\mu)=0$ due to the constraint $\mathrsfso{V} \cdot \mathrsfso{U}=0$, so that $\mu$ is really the moment map for the action of $\mathfrak{sl}_n$. From~(\ref{gaugedSUSY}) one easily finds the SUSY variation of~$\mu$:
\bea
\delta\mu=\upepsilon_2\,\dd\tilde{\mu},\quad\quad\quad \tilde{\mu}:=U\otimes B\,.
\eea
The geometric meaning of $\tilde{\mu}$ is that it is the moment map for the action of the fermionic subgroup $\mathsf{\Pi} (\mathfrak{sl}_n)\mysub \GL(n|n)$ featuring in~(\ref{normalizer}).

\vspace{0.3cm}\noindent
The interacting Lagrangian has the form (the variables $\mathrsfso{U}, \mathrsfso{V}$ were defined in~(\ref{doublets}))
\begin{empheq}[box=\fbox]{align}
\hspace{1em}\vspace{1em} \label{SUSYlagr}
&\mathrsfso{L}_{\CP}=
\mathrsfso{V}\cdot \bar{\mathcal{D}}\mathrsfso{U}+ \bar{\mathcal{U}}\cdot \mathcal{D}\bar{\mathrsfso{V}}
+\vkappa\,\mathrm{Tr}(\mu \bar{\mu})\,,\\ \nonumber
&\textrm{where}\quad\quad \bar{\mathcal{D}} =\bd +i\,\bar{\mathcal{A}}_{\mathrm{super}}\,,\quad\quad \bar{\mathcal{A}}_{\mathrm{super}}=\begin{pmatrix}
      \bar{\mathcal{A}} & 0  \\
      \bar{\mathcal{W}} &  \bar{\mathcal{A}}
   \end{pmatrix}
\end{empheq}
One can view the fermionic variable $\bar{\mathcal{W}}$ either as a Lagrange multiplier imposing the constraint~(\ref{mm2}) or as a gauge field for the local symmetry~(\ref{locsymm}), and $\vkappa$ is the coupling constant. To make sure that the Lagrangian is invariant w.r.t. the SUSY transformations on-shell, we write out the e.o.m.:
\bear
&&V:\quad\bar{\mathrsfso{D}} U+\vkappa\,\bar{\mu}U=0,\quad\quad\quad\quad U:\quad \bar{\mathrsfso{D}} V+\bar{\mathcal{W}} \,B-\vkappa\,V\bar{\mu}=0\,,  \\
&&C:\quad \bar{\mathrsfso{D}} B-\vkappa\,B\bar{\mu}=0,\quad\quad\quad\quad B:\quad \bar{\mathrsfso{D}} C+\bar{\mathcal{W}}\,U+\vkappa\,\bar{\mu}C=0\,. 
\eear
As a result we get the very simple e.o.m. for the moment map\footnote{Here $\bar{\mu}$ is the moment map for the complex conjugate symplectic form, as defined by the kinetic term in~(\ref{SUSYlagr}): $\bar{\mu}=\bar{V}\otimes \bar{U}-\bar{B}\otimes \bar{C}$. Its e.o.m. $\dd \bar{\mu}=\vkappa\,[\bar{\mu}, \mu]$ has an unusual-looking sign due to the fact that the kinetic term in~(\ref{SUSYlagr}) is imaginary, which is a consequence of working in Euclidean signature (in fact, it ensures convergence of the path integral for $\vkappa>0$), see the discussion in~\cite{BykovGN}.}
\bea\label{mueom}
\bd\mu=-\vkappa\, [\bar{\mu}, \mu]\,.
\eea
We can now compute the SUSY variation of the interaction term:
\bea
\delta (\vkappa\,\mathrm{Tr}(\mu \bar{\mu}))=\upepsilon_2\,\vkappa\,\mathrm{Tr}(\dd\tilde{\mu}\cdot\bar{\mu})\sim\,-\upepsilon_2\,\vkappa^2\,\mathrm{Tr}([\mu, \tilde{\mu}]\,\bar{\mu})\,,
\eea
where $\sim$ means `up to integration by parts', and in the final equality we have used the e.o.m.~(\ref{mueom}). An elementary calculation shows that the commutator
\bea\label{mumutcomm}
[\mu, \tilde{\mu}]=(V\cdot U+B\cdot C)\,\tilde{\mu}-(B\cdot U)\,\mu=0
\eea
vanishes as a consequence of the constraints~(\ref{mm1}) and (\ref{mm2}). The action is therefore invariant under SUSY transformations on-shell. In Appendix~\ref{SUSYdetailsapp} we demonstrate that an off-shell-invariant formulation may also be constructed, which is in agreement with the standard SUSY transformations~\cite{WittenBeta, Kapustin}.

\vspace{0.3cm}\noindent
As discussed earlier, the symmetry group of the kinetic term in~(\ref{SUSYlagr}) is $(\CC^\times\!\times\PSL(n))\ltimes \mathsf{\Pi} (\mathfrak{sl}_n)$. The interaction term further breaks this to the $R$-symmetry times the unitary subgroup $\CC^\times\times(\mathbf{PSU}(n)\simeq \mathbf{SU}(n)/\mathbb{Z}_n)$, which is the global symmetry group of the supersymmetric $\CP^{n-1}$-model.

\subsubsection{Conventional formulation.}\label{usualSUSYsec}

To relate the model~(\ref{SUSYlagr}) to the more standard form of the supersymmetric $\CP^{n-1}$-model, first of all one has to eliminate the $V, \bar{V}$-fields, just as in the bosonic case. However even after such elimination the $\G_\triangle$ super gauge symmetry~(\ref{Gtriang}) still remains, and one might wish to pick a gauge to fix it. The commonly used gauge for the scaling $\mathbb{R}^+\mysub \CC^\times$ gauge symmetry is
\bea
\label{spheregauge}
\bar{U}\cdot U=1\,.
\eea

\noindent
After this condition is imposed, one still has the local symmetry~(\ref{locsymm}). The combination $\bar{U}\cdot C$ shifts under this symmetry as $\delta_{\mathrm{loc}}(\bar{U}\cdot C)=\updelta(z, \bar{z})\,\bar{U}\cdot U=\updelta(z, \bar{z})$, so that a simple gauge choice is
\bea
\label{UCortho}
\bar{U}\cdot C=\bar{C}\cdot U=0\,.
\eea

\noindent
This condition may be also seen as a superpartner to~(\ref{spheregauge}) w.r.t. the SUSY transformations~(\ref{gaugedSUSY}) and their left-moving counterparts. As a result, the constraints~(\ref{spheregauge})-(\ref{UCortho}) are supersymmetric. The condition~(\ref{UCortho}), together with the moment map constraint~(\ref{mm2}), $B\cdot U=0$, may be succinctly rephrased as
\bea\label{Diracortho}
\bar{U} \cdot \Theta=0\,,
\eea
where $\Theta$ is the Dirac fermion $\Theta=\begin{pmatrix}C\\ \bar{B}\end{pmatrix}$. It is the constraint~(\ref{Diracortho}) that is most commonly encountered in the literature on the SUSY $\CP^{n-1}$-model. We wish to emphasize that in our approach only the moment map condition~(\ref{mm2}) should be seen as fundamental, whereas~(\ref{UCortho}) is simply a gauge choice. Nevertheless the gauge choice~(\ref{spheregauge})-(\ref{UCortho}) is rather convenient, since in this case the bosonic and fermionic parts in the interaction terms separate: $\mathrm{Tr}(\mu \bar{\mu})=V\cdot \bar{V}-(\bar{C}\cdot C)\, (B\cdot\bar{B})$. Upon integration over the $V, \bar{V}$-variables one obtains the conventional form of the supersymmetric Lagrangian\footnote{In our notation $\gamma_5:=i\,\sigma_1\sigma_2=-\sigma_3$.}:
\bear\label{SUSYlagr2}
&&\mathrsfso{L}_{\CP}={1\over \vkappa}\,\big|\bar{\mathrsfso{D}} U\big|^2+\bar{\Theta} \slashed{\mathrsfso{D}}\Theta-\vkappa\,\left(\bar{\Theta}{1+\gamma_5\over 2}\Theta\right)\left(\bar{\Theta}{1-\gamma_5\over 2}\Theta\right)\\ \nonumber
&&\textrm{where}\quad\quad \bar{U}\cdot U=1,\quad\quad \bar{U} \cdot \Theta=0\,.
\eear
In other words, supersymmetrization involves coupling the bosonic part of the Lagrangian to a chiral Gross-Neveu model, as was noted long ago~\cite{AbdallaCPN}. The present approach based on $\beta\gamma$-systems elucidates the origin of this phenomenon: the bosonic part is itself a chiral Gross-Neveu model.

\vspace{0.3cm}\noindent
We may obtain yet another form of the model by eliminating the auxiliary gauge field $\mathcal{A}, \bar{\mathcal{A}}$. One finds $\mathcal{A}=A_b-i\,\vkappa\,B \cdot C$ and $\bar{\mathcal{A}}=\bar{A}_b+i\,\vkappa\,\bar{C}\cdot\bar{B}$, where $A_b:=-i\, \dd\bar{U}\cdot U$ and $\bar{A}_b:=i\,\bar{U}\cdot \bd U$ are the purely bosonic parts of the connection. A simple rewriting then gives
\bea
\mathrsfso{L}_{\CP}={1\over \vkappa}\,\big|\bar{\mathrsfso{D}}_b U\big|^2+\bar{\Theta} \slashed{\mathrsfso{D}}_b\Theta-{\vkappa\over 4}\left((\bar{\Theta}\Theta)^2-(\bar{\Theta}\sigma_3\Theta)^2+(\bar{\Theta}\sigma_i\Theta)^2\right)\,,
\eea
where $\mathrsfso{D}_b, \bar{\mathrsfso{D}}_b$ are the covariant derivatives w.r.t. the connection $A_b, \bar{A}_b$. Again, one has the additional conditions $\bar{U}\cdot U=1,\, \bar{U} \cdot \Theta=0$. It is this form of the model that one finds in~\cite{AbdallaCPN}, for example.

\section{Anomalies}\label{anomalysec}

In this section we pass to the discussion of potential gauge anomalies in the models introduced in the previous sections. The first important observation is that part of the symmetry we are gauging in models like~(\ref{lagrkin1}) or~(\ref{SUSYlagr}) is actually chiral, and for this reason it is typically subject to anomalies. To see this, note an important difference in chiral symmetry transformations for Minkowski vs. Euclidean signature of the worldsheet. If $\mathsf{G}$ is the compact group of (vectorial) symmetry transformations of the theory, the chiral symmetry group~is
\bear\label{chiral1}
&&\textrm{Minkowski signature:}\quad\quad \mathsf{G}\times \mathsf{G}\\ \label{chiral2}
&&\textrm{Euclidean signature:}\quad\quad\;\;\;\; \mathsf{G}_{\CC}
\eear
This fact was observed as early as in~\cite{Zumino, Mehta}, and it has a bearing on most aspects of the theory related to anomalies. For example, the target space of the corresponding WZNW theory is $\mathsf{G}\simeq {\mathsf{G}\times \mathsf{G}\over \mathsf{G}}$ and $\mathsf{G}_{\CC}\over \mathsf{G}$ respectively. In the case $\mathsf{G}=\mathbf{U}(n)$ the latter is the space of Hermitian positive-definite matrices, and the corresponding WZNW model was thoroughly studied in~\cite{Gawedzki}. We also note that the complexified gauge groups naturally arise in supersymmetric theories, cf.~\cite{Lerche} where the reader will also find a corresponding 4D anomaly cancellation condition.

\vspace{0.3cm}\noindent
Due to the difference between~(\ref{chiral1}) and~(\ref{chiral2}) the condition of anomaly cancellation~\cite{WittenGauged} is suitably modified. Indeed, according to~\cite{WittenGauged} a subgroup $\mathsf{H}$ of the chiral symmetry group may be gauged if the following condition is satisfied: $\mathrm{Tr}_L(T_i\,T_j)=\mathrm{Tr}_R(T_i\,T_j)$, where $L$ and $R$ are the $\mathsf{H}$-representations of the left- and right-handed fields. Since in the case of Euclidean signature the left- and right-handed fields transform in complex conjugate representations of $\mathsf{G}_{\CC}$, one has $(T_i)_R=\bar{(T_i)_L}$. It then follows from the cancellation condition that, as expected, $\mathsf{H}=\mathsf{G}\mysub \mathsf{G}_{\CC}$ may be gauged, since in this case $(T_i)_R=(T_i)_L$. We are interested, however, in gauging complex subgroups of $\mathsf{G}_{\CC}$. This means that if $T_j$ is a generator, so is $i\,T_j$. As a result, one has, two conditions $\mathrm{Tr}_L(T_i\,T_j)=\pm\mathrm{Tr}_L(\bar{T_i}\,\bar{T_j})$, or equivalently $\mathrm{Tr}_L(T_i\,T_j)=0$ for all pairs of generators of the Lie algebra $\mathfrak{g}_{\CC}$. In the setup with target space supersymmetry, i.e. when the global symmetry group is in fact a supergroup, one has both fermions and bosons contributing to the anomaly, and this condition is promoted~to
\begin{empheq}[box=\fbox]{align}
\label{anomcancel}
\quad \mathrm{Str}_{\mathsf{W}}(T_i T_j)=0\,,\quad
\end{empheq}
where $\mathsf{W}$ is the representation of the (left-handed) matter fields. In the examples we encountered earlier (the two phase spaces $\Phis_{\mathrm{min}}$ and $\Phis_{\mathrm{SUSY}}$) the supertrace of the generator vanishes as well, and we expect this to hold in most cases of interest:
\bea\label{anomcancel2}
\mathrm{Str}_{\mathsf{W}}(T_i)=0\,.
\eea
When one of the generators is the identity matrix, the latter condition clearly follows from~(\ref{anomcancel}). It would be interesting to understand the precise relation of the conditions~(\ref{anomcancel})-(\ref{anomcancel2}) to the vanishing of $\beta\gamma$-system anomalies~\cite{WittenBeta, Nekrasov} (i.e. $ch_2(\Ms)=0$ and possibly ${c_1(\Ms)=0}$ for the target space $\Ms$), see also~\cite{MooreAnomaly} and references therein for a general discussion of $\sigma$-model anomalies. In our application to the $\CP^{n-1}$-model the above anomaly cancellation conditions mean, in particular, that the model in the `Hopf fibration' gauge $\bar{U}\cdot U=1$ is equivalent to the model in the `inhomogeneous' gauge $U_n=1$. The relation between these gauges in the purely bosonic model has been studied in~\cite{Rabinovici, BandoKugo}.

\begin{figure}
\centering
\bea\nonumber
\begin{tikzpicture}[thick]                    
                    \path [draw=mygreen1,snake it, line width=2pt]
    (4,0) -- (5,0) ;
    \path [draw=mygreen1,snake it, line width=2pt]
    (7,0) -- (8,0) ;
    \draw[blue!50, line width=2pt, rounded corners] (6,1) to [out=180,in=90] (5,0);
        \draw[-stealth, blue!50, line width=2pt, rounded corners] (7,0) to [out=90,in=0] (5.9,1);
            \draw[blue!50, line width=2pt, rounded corners] (6,-1) to [out=0,in=-90] (7, 0);
                    \draw[-stealth, blue!50, line width=2pt, rounded corners] (5,0) to [out=-90,in=180] (6.1,-1);
                    \node at (4.3,0.4) {$\mathcal{A}$};
                    \node at (7.7,0.4) {$\mathcal{A}$};
\end{tikzpicture}
\eea
\caption{The anomaly cancellation condition~(\ref{anomcancel}) corresponds to the vanishing of the sum of such diagrams. The matter fields propagating in the loops involve both bosonic and fermionic fields.} \label{fig1}
\end{figure}
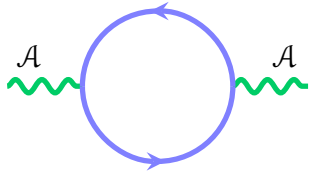

\vspace{0.3cm}\noindent
The above conditions in fact imply that the effective action is independent of the gauge fields. Indeed, at quadratic order in the gauge field the diagram shown in Fig.~\ref{fig1} is proportional to the trace in~(\ref{anomcancel}) and therefore vanishes. As for higher orders, this is demonstrated at the level of the Feynman diagrams in Appendix~\ref{determinantapp} and is a consequence of the fact that all diagrams contributing to the effective action are in fact proportional to the quadratic form given by the supertrace $\mathrm{Str}_{\mathsf{W}}$.  The same conclusion may be reached by performing a calculation in the spirit of~\cite{PW}, which shows that the determinant of the Dirac operator in an external gauge field is proportional to $e^{c_{\mathsf{W}}\cdot\mathsf{S}_{\mathrm{WZNW}}}$, where $\mathsf{S}_{\mathrm{WZNW}}$ is the WZNW action (whose fields are expressed in terms of the gauge field of the original model), and $c_{\mathsf{W}}$ is a proportionality constant characterizing the quadratic form given by the (super)-trace: $\mathrm{Tr}_{\mathsf{W}}(T_i T_j)=c_{\mathsf{W}}\cdot\mathrm{Tr}_{\mathsf{W}_0}(T_i T_j)$ (here $\mathsf{W}_0$ is some reference representation). This calculation is reviewed in~\cite{Nair} (see~\cite{Nair1, Nair2} for the original presentation). For $c_{\mathsf{W}}=0$ the determinant is a constant, independent of the gauge field.

\vspace{0.3cm}\noindent
Independence of the effective action of the gauge field is indeed necessary for the following reason. Under the complexified infinitesimal gauge transformations the components of the gauge field transform as $\delta\mathcal{A}=\dd \chi$, $\delta\bar{\mathcal{A}}=\bd \bar{\chi}$. Apart from total derivative terms, in the infinite-volume theory there is no way to build an invariant combination of $\mathcal{A}, \bar{\mathcal{A}}$. In finite volume, or on a torus, one could have gauge-invariant holonomies $\int\limits_\gamma \,\mathcal{A}$ over cycles $\gamma$ on the worldsheet. We pass to a discussion of this fact on the simpler example of a quantum mechanical model at finite temperature (i.e. on a circle $S^1$).

\section{Quantum mechanical model}\label{QMsec}

In the previous section we argued that, for a non-anomalous model, the infinite-volume partition function $\mathsf{Z}_{\infty}(\mathcal{A}, \bar{\mathcal{A}})$ of the matter fields has to be independent of the gauge fields $\mathcal{A}, \bar{\mathcal{A}}$. Let us now explain this statement from the point of view of the quantum-mechanical reduction of the model. We start by considering the quantum mechanical partition function $\mathsf{Z}_{\upbeta}$ for a $\CP^{n-1}$-model on a Euclidean time circle of circumference~$\upbeta$:
\bear\label{QMpartfunc}
&&\mathsf{Z}_{\upbeta}(\mathcal{A}_0, \bar{\mathcal{A}}_0)=\int\,e^{-\int\limits_0^\upbeta\,dt\,\mathrsfso{L}_{\CP}}\,\prod\limits_t\,dU d\bar{U}dVd\bar{V},\\
&&\mathrsfso{L}_{\CP}= (V\cdot \bar{\mathrsfso{D}} U+\bar{U}\cdot\mathrsfso{D} \bar{V})+\vkappa\,(\bar{U}\cdot U)(\bar{V}\cdot V)
\eear
The notation $\mathsf{Z}_{\upbeta}(\mathcal{A}_0, \bar{\mathcal{A}}_0)$ indicates that the partition function depends really only on the zero-modes $\mathcal{A}_0:={1\over \upbeta}\,\int\,dt\,\mathcal{A}$ of the gauge fields, which are the only gauge invariants for the $\CC^\times$ gauge transformations (these are the same as the holonomies $\int\limits_\gamma \,\mathcal{A}$ mentioned at the end of the previous section). A priori one assumes periodic boundary conditions in the path integral~(\ref{QMpartfunc}): $U(\upbeta)=U(0), V(\upbeta)=V(0)$. One can get rid of the zero modes of the gauge fields at the expense of imposing twisted boundary conditions on the fields. Indeed, the change of variables $U\to e^{i\,\mathcal{A}_0\,t} U, V\to e^{-i\,\mathcal{A}_0\,t} V$ eliminates the gauge field, but leads to the twisted boundary conditions
\bea
U(\upbeta)=h\circ U(0), \quad\quad V(\upbeta)=h^{-1}\circ V(0)\,,\quad\quad h\in \mathbf{H}\,,
\eea
where $\mathbf{H}$ is the original complex  gauge group ($\mathbf{H}=\CC^\times$ in the case of the $\CP^{n-1}$-model, although the discussion here applies more generally). To compute the full partition function one now has to integrate $\mathsf{Z}_{\upbeta}(\mathcal{A}_0, \bar{\mathcal{A}}_0)$ over the zero modes $\mathcal{A}_0, \bar{\mathcal{A}}_0$. Clearly, this is the same as integrating over the twists $h$. Recalling the definition of the partition function as the trace of the statistical operator $e^{-\upbeta \mathsf{H}}$, we get\footnote{Here we define $| h \bar{h} \Psi\rangle:=(h\otimes \bar{h}) | \Psi\rangle $. Writing this, we treat the Hilbert space as being the tensor product of the holomorphic and anti-holomorphic spaces.}
\bea
\mathsf{Z}_{\upbeta}:=\int\,dh\,d\bar{h}\, \mathsf{Z}_{\upbeta}(\mathcal{A}_0, \bar{\mathcal{A}}_0)=\int\,dh\,d\bar{h}\,d\Psi\,\langle  \Psi| e^{-\upbeta \mathsf{H}}|h \bar{h} \Psi\rangle\,.
\eea
Since $\int\,dh\,h=\mathsf{\Pi_{\mathbf{H}}}$ is the projector on the invariants of $\mathbf{H}$, we find
\bea
\mathsf{Z}_{\upbeta}=\mathrm{Tr}(\mathsf{\Pi}_{\mathbf{H}} e^{-\upbeta \mathsf{H}})=\mathrm{Tr}_{\mathrm{\mathbf{H}-inv}}(e^{-\upbeta \mathsf{H}})\,.
\eea
In other words, the role of the gauge fields in `finite volume' (i.e. for finite values of $\upbeta$ in this case) is to restrict the Hilbert space to a subsector invariant under the action of $\mathbf{H}$. The infinite volume limit $\upbeta\to\infty$ corresponds to picking the ground state in the spectrum. The fact that the partition function is independent of the gauge field in this limit means that the ground state is symmetric w.r.t. the symmetry group and is therefore shared by all the models with different gauge fields.

\section{Grassmannian and quiver generalizations}\label{quiversec}

The setup of the $\CP^{n-1}$-model can be straightforwardly generalized to the case of Grassmannians $\mathsf{G}(m, n)$. The first step in doing this is realising that the configuration space $\CC^{n|n}$ may be thought of as $\mathrm{Hom}(\CC^{1|1}, \CC^n)$. The natural generalization is then to take $\mathrm{Hom}(\CC^{m|m}, \CC^n)$ and consider the quotient
\bea
\Ms_{\mathsf{G}(m, n)}:=\mathrm{Hom}(\CC^{m|m}, \CC^n) / \G_{\triangle}\,, 
\eea
where as $\G_{\triangle}$ one has to take the relevant subgroup of $\SL(m|m)$:
\bea\label{GtriangGrass}
\G_{\triangle}:=\left\{\quad g\in \SL(m|m)\,:\quad g=\begin{pmatrix} 
      \uplambda & 0  \\
      \upxi &  \uplambda
   \end{pmatrix},\quad\quad \uplambda \in \GL(m)\quad \right\}\,,
\eea
where $\upxi$ is now an $m\times m$ fermionic matrix. The phase space with stable subset $\Phis^{(st)}=\mathsf{T^\ast} \Ms_{\mathsf{G}(m, n)}$ is described by the following elementary quiver:
\bea
\begin{tikzpicture}[
baseline=-\the\dimexpr\fontdimen22\textfont2\relax,scale=1]
\begin{scope}[very thick,decoration={
    markings,
    mark=at position 0.4 with {\arrow{>}}}
    ] 
\draw[postaction={decorate}] ([yshift=-2pt,xshift=0pt]0,0) --  node [below, yshift=0pt] {\footnotesize $\mathrsfso{U}$} ([yshift=-2pt,xshift=0pt]2,0);
\end{scope}
\begin{scope}[very thick,decoration={
    markings,
    mark=at position 0.6 with {\arrow{<}}}
    ] 
\draw[postaction={decorate}] ([yshift=2pt,xshift=0pt]0,0) --  node[above, yshift=0pt] {{\footnotesize $ \mathrsfso{V}$}} ([yshift=2pt,xshift=0pt]2,0);
\end{scope}
\filldraw[black!50] (1.85,-0.15) rectangle ++(8pt,8pt);
\filldraw[black!50] (0,0) circle (4pt); 
\node at (2.5,0.5) {$\CC^n$};
\node at (-0.5,-0.5) { $\CC^{m|m}$};
\end{tikzpicture}
\eea
Here $\mathrsfso{U}\in \mathrm{Hom}(\CC^{m|m}, \CC^n)$ and $\mathrsfso{V} \in \mathrm{Hom}(\CC^n, \CC^{m|m})$.

\vspace{0.3cm}\noindent
Instead of taking $\uplambda \in \GL(m)$ one could as well take $\uplambda$ in some parabolic subgroup $P\mysub \GL(m)$. This will give rise to $\sigma$-models with flag manifold target spaces~\cite{BykovFlag1, BykovFlag2, BykovFlag3, BykovGLSM1, BykovGLSM2, BykovAnom, CYa, BykovNilp, BykovGN}, coupled to fermions. In general, however, such models will no longer have worldsheet~SUSY.

\vspace{0.3cm}\noindent
This discussion suggests a further generalization. Suppose we have a super phase space~$\mathsf{\Phi}$, which is a complex symplectic (quiver) supervariety. There is a gauge (super)-group $\G_{\textrm{gauge}}$ acting in the nodes of the quiver, and matter fields $\mathrsfso{U}\in \mathsf{W}$, $\mathrsfso{V}\in \mathsf{W}^{\vee}$ are in representations $\mathsf{W}\oplus \mathsf{W}^{\vee}$ of $\G_{\textrm{gauge}}$. We assume that the quiver is `doubled', meaning that every representation arises together with its dual (Nakajima quivers have this property~\cite{Nakajima, NakajimaReview}). Apart from the gauge nodes, the quiver will typically have some global nodes with an action of a global symmetry complex group~$\G_{\textrm{global}}$. We can therefore define the complex moment map $\mu$ for the action of $\G_{\textrm{global}}\circlearrowright\mathsf{\Phi}$. In this setup one can, quite naturally, define the following Lagrangian:
 
\vspace{-0.2cm}
\begin{empheq}[box=\fbox]{alignat=3}
\hspace{1em}\vspace{1em} \label{SUSYlagrGen}
&\quad \mathrsfso{L}=\left(\mathrsfso{V}\cdot \bar{\mathcal{D}} \mathrsfso{U}+\bar{\mathrsfso{U}}\cdot \mathcal{D} \bar{\mathrsfso{V}}\right)+\vkappa\,\mathrm{Tr}(\mu \,\bar{\mu})\,.\quad
\end{empheq}
We encountered various special cases of this model earlier in this paper, cf.~(\ref{SUSYlagr}). The kinetic term in the Lagrangian corresponds to the $\beta\gamma$-systems -- it is a pull-back of the canonical Liouville one-form corresponding to the complex symplectic form of the quiver. The second term provides a coupling between the holomorphic and (anti)-holomorphic $\beta\gamma$-systems and comes with an arbitrary coefficient $\vkappa$ that should be seen as a coupling constant (in the $\sigma$-model setup this is the inverse squared radius of the target space). In Appendix~\ref{symplapp} we show directly that the moment map $\mu$ satisfies the e.o.m.~(\ref{mueom}) of the principal chiral model in this more general situation as well.

\vspace{0.3cm}\noindent
As we discussed in the previous sections, one also needs to impose the chiral anomaly cancellation conditions that in the general setup have the form
\bea\label{anomcancelGen}
\mathrm{Str}_{\mathsf{W}}(T_a T_b)=0\,,\quad \textrm{where} \quad  T_a, T_b\in \mathfrak{g}_{\textrm{gauge}}\,.
\eea

\vspace{-0.2cm}\noindent
As mentioned earlier, we expect that in most cases $\mathrm{Str}_{\mathsf{W}}(T_a)=0$ holds as well. It is tempting to conjecture that the Lagrangian~(\ref{SUSYlagrGen}), supplemented with the conditions~(\ref{anomcancelGen}), defines a quantum integrable model. All the models described earlier in this paper ($\CP^{n-1}$, Grassmannian, flag manifold $\sigma$-models with fermions) are particular examples of this system. We leave further clarification of these issues for the future.

\section{Conclusion}

In the present paper we continued the study of integrable $\sigma$-models with complex homogeneous target spaces~\cite{BykovFlag3}, based on their formulation as gauged bosonic (or mixed bosonic/fermionic) Gross-Neveu models proposed in~\cite{BykovGN}. The main emphasis was on the fermionic generalizations of the well-known $\CP^{n-1}$-model, although the discussion can be generalized to a wide class of `quiver supervarieties'. Our main finding is that all such models may be defined in a canonical way in terms of a target space supervariety (the phase space $\Phis$ or configuration space $\Ms$). The cancellation of chiral gauge anomalies that might be present in such models has been formulated as a simple constraint on these varieties. We conjectured that these chiral anomalies underlie the anomalies in the Yangian charges of the purely bosonic models: in both cases inclusion of fermions cancels the anomalies. As a by-product, we have developed a new method for deriving worldsheet supersymmetric $\sigma$-models by starting from models with target space supersymmetry and gauging part of their symmetry supergroups.

\vspace{0.3cm}\noindent
\textbf{Acknowledgments.} I would like to thank A.~A.~Slavnov for support and D.~L\"ust for reading the manuscript and useful remarks and suggestions.

\vspace{1.5cm}\noindent
{\large \textsf{\emph{Appendix}}}
\begin{appendices}

\vspace{-0.4cm}\noindent
\section{Details on supersymmetry transformations}\label{SUSYdetailsapp}

In section \ref{SUSYsec} we showed that the Lagrangian~(\ref{SUSYlagr}) that describes the target space supersymplectic quotient is in fact invariant under worldsheet supersymmetry transformations on-shell. This means that the variation of the Lagrangian is proportional to the e.o.m. (up to full derivative terms that we drop). In such cases there is a simple tool to make the invariance off-shell. Indeed, suppose that the Lagrangian depends on some generalized coordinates $q^j$, so that the variation of the action gives the e.o.m. $E_j=0$, i.e. $\Delta\mathsf{ S}=\int\,d^2z\,\Delta q^j\,E_j$ (we use the symbol $\Delta$ for the variation to distinguish it from the variation $\delta$ w.r.t. some symmetry). Suppose there is a symmetry $\delta$ such that the variation of the action is proportional to the e.o.m.: $\delta\mathsf{ S}=\int\,d^2z\,\delta V^j\,E_j$, where $\delta V^j$ is a vector field in field space. As a result, the combined variation $\hat{\delta}=\delta+\Delta$, where in the second term we take $\Delta q^j=-\delta V^j$, annihilates the action: $\hat{\delta}\mathsf{S}=0$. To summarize, the `off-shell' variation of the fields takes the form
\bea
\hat{\delta} q^j=\delta q^j-\delta V^j\,.
\eea
In our applications to supersymmetry $\delta q^j$ are the `on-shell' transformations~(\ref{gaugedSUSY}). The variation of the action is (here we use~(\ref{mumutcomm}))
\bear\label{actvar}
&\delta \mathsf{S}=\int\,d^2z\,\upepsilon_2\, \left[F_{z\bar{z}}\,(B \cdot U)-\vkappa\,\mathrm{Tr}(\tilde{\mu}\,(\dd\bar{\mu}+\vkappa [\mu, \bar{\mu}]))+\right.\\ \nonumber 
&\left.-\vkappa^2 (V\cdot U+B\cdot C)\,\mathrm{Tr}(\tilde{\mu}\bar{\mu})+\vkappa^2 (B\cdot U)\,\mathrm{Tr}(\mu\bar{\mu})
\right]
\eear
The coefficients in front of $V\cdot U+B\cdot C$ and $B\cdot U$ may be reabsorbed in the shifts of the gauge fields $\mathcal{A}$ and $\bar{\mathcal{W}}$:
\bear\label{var1}
&&\hat{\delta} (i\bar{\mathcal{A}})=\upepsilon_2\,\vkappa^2\,\mathrm{Tr}(\tilde{\mu}\bar{\mu})\\
&&\hat{\delta} (i \bar{\mathcal{W}})=\upepsilon_2\,(F_{z\bar{z}}+\vkappa^2\,\mathrm{Tr}(\mu\bar{\mu}))
\eear
The coefficient in front of the l.h.s. of the e.o.m. $\dd\bar{\mu}+\vkappa [\mu, \bar{\mu}]$ in~(\ref{actvar}) may be reabsorbed in a $\mathsf{G}_{\CC}$-transformation of the anti-holomorphic fields $\bar{U}, \bar{V}, \bar{B}, \bar{C}$ with parameter $\upepsilon_2\,\vkappa\,\tilde{\mu}$ (no compensating transformation for $U, V, B, C$ is needed). Recalling that $\tilde{\mu}=U\otimes B$, we get
\bear\label{var7}
&&\hat{\delta}\bar{U}=\bar{U}\circ (\upepsilon_2\,\vkappa\,\tilde{\mu})=\upepsilon_2\,\vkappa\,(\bar{U} \cdot U)\, B\\
&&\hat{\delta}\bar{C}= \bar{C}\circ( \upepsilon_2\,\vkappa\,\tilde{\mu})=-\upepsilon_2\,\vkappa\,(\bar{C}\cdot U)\, B\\
&&\hat{\delta}\bar{V}=-\upepsilon_2\,\vkappa\, \tilde{\mu}\circ\bar{V}=-\upepsilon_2\,\vkappa\,(B\cdot \bar{V})\, U\\ \label{var10}
&&\hat{\delta}\bar{B}=-\upepsilon_2\,\vkappa\, \tilde{\mu}\circ\bar{B}=-\upepsilon_2\,\vkappa\, (B\cdot\bar{B})\,U
\eear
Equations~(\ref{var1})-(\ref{var10}) provide a complete set of off-shell SUSY transformations in the formalism with auxiliary gauge fields. 

\vspace{0.3cm}\noindent
To compare with the standard formulations of SUSY gauged linear $\sigma$-models let us choose the gauge discussed in section~\ref{usualSUSYsec}:
\bea\label{gaugeagain}
\bar{U}\cdot U=1, \quad\quad \bar{C}\cdot U=\bar{U}\cdot C=0\,.
\eea
The supersymmetry variations are $Q_+(\bar{U}\cdot U-1)=\bar{U}\cdot C=0$ and $Q_+(\bar{C}\cdot U)=-\bar{C}\cdot C\neq 0$, therefore in order to maintain supersymmetry after gauge fixing we have to perform a compensating fermionic gauge transformation with parameter $\bar{C}\cdot C$. In other words, $\delta \bar{C}=\upepsilon_1\,(\bar{C}\cdot C)\,\bar{U}$.

\vspace{0.3cm}\noindent
Besides, in the standard formulation there are no $V$-fields (since they have been integrated out), so we will also eliminate them using their e.o.m., which gives in our gauge
\bea
\mathrsfso{D} \bar{U}-\vkappa\,\bar{U}\mu=\mathrsfso{D} \bar{U}-\vkappa\,V=0
\eea
Combining the expressions~(\ref{var7})-(\ref{var10}) and~(\ref{gaugedSUSY}) we obtain the final transformation laws for the standard variables (i.e. all variables except $V, \bar{V}$):
\bea
\begin{aligned}[c]
&\hat{\delta} U=\upepsilon_1\,C\,,\\
&\hat{\delta} B=-{\upepsilon_1\over \vkappa}\,\mathrsfso{D} \bar{U}\,,\\
&\hat{\delta} C=-\upepsilon_2\,\mathrsfso{D} U\,,
\end{aligned}
\qquad\qquad\qquad
\begin{aligned}[c]
&\hat{\delta}\bar{U}=\upepsilon_2\,\vkappa\, B\,,\\
&\hat{\delta}\bar{B}=-\upepsilon_2\,\vkappa\, (B\cdot\bar{B})\,U\,,\\
&\hat{\delta}\bar{C}=\upepsilon_1\,(\bar{C}\cdot C)\,\bar{U}\,.
\end{aligned}
\eea
One can check directly that the Lagrangian~(\ref{SUSYlagr2}) is invariant, up to a total derivative, w.r.t. these transformations, which are the standard $(0, 2)$ SUSY transformations for a gauged linear $\sigma$-model after elimination of auxiliary fields, cf.~\cite[Chapter 15]{Mirror} (in~\cite{WittenAB, WittenBeta} one finds analogous transformations for the nonlinear form of the $\sigma$-model). We see that these more complicated transformations arise from the elementary ones~(\ref{gaugedSUSY}) upon going off-shell and fixing the complex gauge symmetry using the gauge conditions~(\ref{gaugeagain}). The supersymmetry algebra still closes only on-shell, as we are not using auxiliary fields. The Lagrangian~(\ref{SUSYlagr}) admits an additional $(2, 0)$ SUSY acting primarily on the anti-holomorphic fields $\bar{U}, \bar{B}, \bar{C}$, with compensating transformations for $U, B, C$, and as a result the supersymmetry is extended to $(2, 2)$, as expected for a K\"ahler target space.

 \section{Contributions to the superdeterminant}\label{determinantapp}

\begin{figure}
\centering
\bea\nonumber
\begin{tikzpicture}[thick]
\begin{scope}[very thick,decoration={
    markings,
    mark=at position 0.6 with {\arrow{>}}}
    ] 

  \path [draw=mygreen1,snake it, line width=2pt]
    (2.3,1.5) -- (1.7,0.7) ;
      \path [draw=mygreen1,snake it, line width=2pt]
    (2.3,-1.5) -- (1.7,-0.7) ;
    \path [draw=mygreen1,snake it, line width=2pt]
    (-0.6,0) -- (0.4,0) ;
         \draw[postaction={decorate}, blue!50, line width=2pt, rounded corners] (1.7,0.7) to [out=140,in=90] (0.4,0);
            \draw[postaction={decorate}, blue!50, line width=2pt, rounded corners] (1.7,-0.7) to [out=40,in=-40] (1.7,0.7);
                    \draw[postaction={decorate}, blue!50, line width=2pt, rounded corners] (0.4,0) to [out=-90,in=-140] (1.7,-0.7);
                    \node at (-1.2,0) {$\mathcal{A}(z_1)$};
                    \node at (3.0,1.4) {$\mathcal{A}(z_2)$};
                   \node at (3.0,-1.4) {$\mathcal{A}(z_3)$};
                    
                  \path [draw=mygreen1,snake it, line width=2pt]
    (8.3,1.5) -- (7.7,0.7) ;
    \path [draw=mygreen1,snake it, line width=2pt]
    (6.1,1.5) -- (6.7,0.7) ;
        \path [draw=mygreen1,snake it, line width=2pt]
    (6.1,-1.5) -- (6.7,-0.7) ;
      \path [draw=mygreen1,snake it, line width=2pt]
    (8.3,-1.5) -- (7.7,-0.7) ;
         \draw[postaction={decorate}, blue!50, line width=2pt, rounded corners] (7.7,0.7) to [out=150,in=25] (6.7,0.7);
            \draw[postaction={decorate}, blue!50, line width=2pt, rounded corners] (7.7,-0.7) to [out=40,in=-40] (7.7,0.7);
                    \draw[postaction={decorate}, blue!50, line width=2pt, rounded corners] (6.7,-0.7) to [out=-25,in=-150] (7.7,-0.7);
          \draw[postaction={decorate}, blue!50, line width=2pt, rounded corners] (6.7,0.7) to [out=-155,in=155] (6.7,-0.7);

                    \node at (5.5,1.4) {$\mathcal{A}{(z_1)}$};
               \node at (9.0,1.4) {$\mathcal{A}{(z_2)}$};
            \node at (9.0,-1.4) {$\mathcal{A}{(z_3)}$};
             \node at (5.5,-1.4) {$\mathcal{A}{(z_4)}$};
\end{scope}
\end{tikzpicture}
\eea
\caption{The diagrams that formally contribute to the effective action.} \label{fig2}
\end{figure}
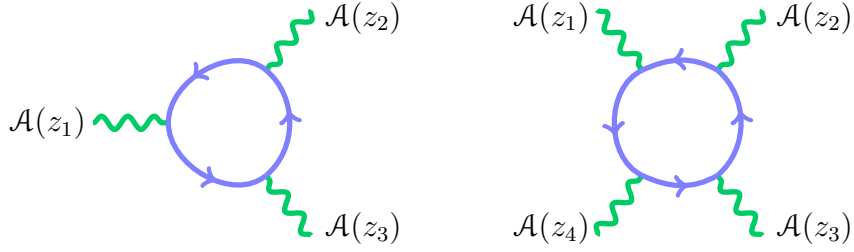
 
In section~\ref{anomalysec} we imposed the anomaly cancellation conditions~(\ref{anomcancel}) on the generators of the gauge superalgebra. These conditions ensure that the sum of diagrams shown in Fig.~\ref{fig1} (with various fields propagating in the loop) vanishes. In fact, all remaining contributions to the `effective action' of the gauge fields $\mathcal{A}, \bar{\mathcal{A}}$ are also proportional to the quadratic form given by the supertrace $\mathrm{Str}_{\mathsf{W}}$ and therefore vanish as well. Let us see how it works on the example of the cubic and quartic vertices shown in Fig.~\ref{fig2}. To this end we introduce the function $G(z_1, \ldots, z_N):=\prod\limits_{j=1}^N\,\frac{1}{z_j-z_{j+1}}$ that involves the product of propagators in the corresponding diagrams~($z_{N+1}=z_1$), and the contribution to the effective action is
\bea
W_N:=\int\,\prod\limits_{j=1}^N\,dz_j\,G(z_1, \ldots, z_N)\,\mathrm{Str}(\mathcal{A}(z_1)\cdots \mathcal{A}(z_N))\,.
\eea
We start from the cubic vertex, $N=3$. Since $G(z_1, z_3, z_2)=-G(z_1, z_2, z_3)$, we have $W_3=\int\,\prod\limits_{j=1}^3\,dz_j\,G(z_1, z_2, z_3)\,{1\over 2}\,\mathrm{Str}(\mathcal{A}(z_1) [\mathcal{A}(z_2),  \mathcal{A}(z_3)])=0$ due to~(\ref{anomcancel}).

\vspace{0.3cm}\noindent
The analysis of the quartic vertex ($N=4$) is slightly more involved. Here we may use the elementary `propagator rearrangement' identity ${1\over (z_3-z_4)(z_4-z_1)}={1\over z_3-z_1}\left({1\over z_3-z_4}+{1\over z_4-z_1}\right)$ to write $G_4(z_1, z_2, z_3, z_4)=\frac{G_3(z_1, z_2, z_3)}{z_3-z_4}+\frac{G_3(z_1, z_2, z_3)}{z_4-z_1}$. Due to the antisymmetry of $G_3$ w.r.t. any pair of arguments, the first and second terms are antisymmetric w.r.t. $z_1\leftrightarrow z_2$ and $z_2\leftrightarrow z_3$ respectively. In the first term in the integrand we therefore replace $\mathrm{Str}(\mathcal{A}(z_1)\mathcal{A}(z_2)\mathcal{A}(z_3) \mathcal{A}(z_4))\to {1\over 2} \mathrm{Str}([\mathcal{A}(z_1), \mathcal{A}(z_2)]\mathcal{A}(z_3) \mathcal{A}(z_4))$, and in the second term we perform an analogous replacement for $\mathcal{A}(z_2) \mathcal{A}(z_3)$. Finally, in the integral of the second term we make a cyclic change of variables $z_1\to z_4\to z_3 \to z_2 \to z_1$ to arrive at an expression skew-symmetric w.r.t. $z_3\leftrightarrow z_4$. As a result,
\bea
W_4:=\int\,\prod\limits_{j=1}^4\,dz_j\,\frac{G_3(z_1, z_2, z_3)+G_3(z_1, z_2, z_4)}{4\,(z_3-z_4)}\,\mathrm{Str}([\mathcal{A}(z_1), \mathcal{A}(z_2)][\mathcal{A}(z_3), \mathcal{A}(z_4)])
\eea
This demonstrates that $W_4$ is as well proportional to the quadratic form given by $\mathrm{Str}$. All the higher contributions $W_N$ (as well as diagrams involving interaction vertices proportional to $\vkappa$) may be analyzed in a similar fashion. 

\section{Moment map evolution for arbitrary complex symplectic manifolds}\label{symplapp}

Let $\mathsf{M}$ be a complex manifold with coordinates $q^i$. In this case $\mathsf{\Phi:=T^\ast M}$ is a complex symplectic manifold with symplectic form $\Omega=\sum\limits_{i=1}^{\mathrm{dim}\,\Ms}\,dp_i\wedge dq^i$. Suppose $\Ms$ is endowed with an action of a complex Lie group $\G$, defined by vector fields $V_a(q)$ ($a=1, \ldots , \mathrm{dim}(\G)$) forming the Lie algebra $\mathfrak{g}$ of $\G$: $[V_a, V_b]=f_{ab}^c\,V_c$. We will assume that $\mathfrak{g}$ admits a non-degenerate $ad$-invariant quadratic form, so that the structure constants $f_{abc}$ are totally skew-symmetric. The moment maps for the action of $\G$ are 
\bea\label{Cmomap}
\mu_a=\sum\limits_{i=1}^{\mathrm{dim}\,\Ms} \,(V_a)^i\,p_i\,,\quad\quad a=1, \ldots , \mathrm{dim}(\G)\,.
\eea
Consider the Lagrangian
\bea\label{lagrsympl}
\mathrsfso{L}=\left(\sum\limits_{i=1}^{\mathrm{dim}\,\Ms}\,p_i \bd q^i-\mathrm{c.c.}\right)+\vkappa\,\sum\limits_{a=1}^{\mathrm{dim}\, G} \big|\mu_a\big|^2
\eea
The e.o.m. for the holomorphic coordinates $(q, p)$ are
\bear
&&\bd q^i+\vkappa\,\sum\limits_{a=1}^{\mathrm{dim}\, G} (V_a)^i \bar{\mu}_a=0\\
&&\bd p_i-\vkappa\,\sum\limits_{a=1}^{\mathrm{dim}\, G} \left(\sum\limits_{j=1}^{\mathrm{dim}\,\Ms} \,\dd_i(V_a)^j\,p_j\right) \bar{\mu}_a=0 \,.
\eear
This induces the following equations for the evolution of the moment maps:
\bear
&&\bd \mu_a=\sum\limits_{i, j} \,\bd q^j\dd_j(V_a)^i\,p_i+\sum\limits_i \,(V_a)^i\,\bd p_i=\\
&&=-\vkappa\,\sum\limits_{i, j, b} \,(V_b)^j\dd_j(V_a)^i\,p_i  \,\bar{\mu}_b+\vkappa\,\sum\limits_{i, j, b} \,(V_a)^i \,\dd_i(V_b)^j\,p_j\, \bar{\mu}_b=\\
&&=\vkappa \,\sum\limits_{j, b}\;[V_a, V_b]^j\,p_j\,\bar{\mu}_b=\vkappa \,\sum\limits_{b, c}\,f_{ab}^c\,\mu_c\,\bar{\mu}_b\,.
\eear
In other words, we have again arrived at the equation
\bea\label{mueq0}
\bd \mu=\vkappa\,[\bar{\mu}, \mu]\,.
\eea
Notice that in the derivation we have not assumed that $\Ms$ is a homogeneous space. Equation~(\ref{mueq0}) is the e.o.m. of the principal chiral model (for more details see~\cite{BykovNilp}), which is the first hint that the model~(\ref{lagrsympl}) might be integrable.

\end{appendices}

\setstretch{0.8}
\setlength\bibitemsep{5pt}
\printbibliography

\end{document}